\documentclass[twocolumn,english,aps,prl]{revtex4-1}
\usepackage[T1]{fontenc}
\usepackage[latin9]{inputenc}
\usepackage{color}
\usepackage{amsmath}
\usepackage{amssymb}
\usepackage{graphicx}
\usepackage{esint}

\usepackage{babel}

\definecolor{cobalt}{RGB}{0, 0,171}
\begin{document}

\title{Effect of nonmagnetic  impurities on $s_\pm$ superconductivity in the presence of incipient bands}

\author{X. Chen$^1$, V. Mishra$^2$, S. Maiti$^{1}$,
and P. J. Hirschfeld$^{1}$}

\affiliation{~$^{1}$Department of Physics, University of Florida, Gainesville,
FL-32611, USA.}
\affiliation{~$^{2}$Joint Institute of Computational Sciences, University of Tennessee, Knoxville, TN-37996, USA.}

\date{\today}
\begin{abstract}
Several Fe chalcogenide superconductors without hole pockets at the Fermi level display high temperature
superconductivity, in apparent contradiction to naive spin fluctuation pairing arguments.  Recently, scanning tunneling microscopy measurements have measured the influence of impurities on some of these materials, 
and claimed that non-magnetic impurities do not create in-gap states, leading to the conclusion that the gap must be $s_{++}$, i.e. conventional $s$ wave with no gap sign change.   Here we critique this argument, and give various ways sign-changing gaps can be consistent with the absence of such bound states.   In particular, we calculate the bound states for an $s_\pm$ system with a hole pocket below the Fermi level, and show that the nonmagnetic impurity bound state energy generically tracks the  gap edge $E_{min}$ in the system, thereby rendering it unobservable.  A failure to observe  a bound state in the case of a nonmagnetic impurity can therefore not be used as  an argument to exclude sign-changing pairing states.
\end{abstract}
\maketitle
\emph{Introduction} ---
Superconductivity in the Fe-based superconductors\cite{ChubukovHirschfeldPhysToday}  is thought to be controlled by local  Coulomb interactions that give rise to repulsive effective interactions between band electrons\cite{HirschfeldCRAS,Chubukov_bookchapter,Qimiao_Review2016}.  In contrast to other materials classes where similar unconventional pairing mechanisms are at work, the Fe-based systems have a Fermi surface consisting of several small pockets around high symmetry points in the Brillouin zone.   Pair scattering between $\Gamma$-centered hole pockets and $M$-centered electron pockets,  was proposed early on as a plausible mechanism, with interpocket repulsion enhanced over intrapocket by electronic spin fluctuations.     Somewhat later, new materials subfamilies were discovered (chiefly the FeSe intercalates\cite{AlkaliIntercalate,AmmoniaIntercalate,LiHydroxideIntercalate} and monolayer FeSe on SrTiO$_3$ (STO)\cite{Xue_ChinPhysLett}),
where doubt was cast on this particular pairing mechanism because of angle-resolved photoemission (ARPES) measurements that showed that the  hole bands were not present at the Fermi surface, but instead had band maxima
50-100meV below (``incipient" bands).

 The remaining electron pockets can also support sign-changing pairing states driven by spin fluctuations, either in the $d$-wave channel\cite{FWangdwave,Maierdwave} or in the so-called ``bonding-antibonding $s$-wave" channel\cite{Mazin_b-a_swave,HKMROPP2011}, depending on the degree of hybridization between the electron pockets\cite{Khodas}.    In addition, ``conventional" sign-changing $s_\pm$ states with gaps of  different signs  on the
 incipient hole band and the electron Fermi pockets are possible\cite{Innocenti,Koshelev,ChenIncipient}.  In one-band systems, when electronic states are moved off the Fermi level, the pairing is rapidly suppressed.  In multiband systems, however, Fermi surface based interactions can stabilize pairing, in which case incipient bands may strongly enhance pairing, and exhibit large gaps\cite{ChenIncipient}.  In addition, it has recently been shown that even without such robust Fermi surface-based interactions, interband interactions with the incipient band alone can create high-$T_c$ superconductivity if one is close to a magnetic instability\cite{LinscheidIncipientVsf,Mishra_bilayer}.

 Recently,  STM studies of the high-quality  surfaces of FeSe intercalates\cite{Fan_etal_NatPhys2015}  and monolayers\cite{Yan_etal_aXv2015} have claimed to rule out sign changing superconducting pair states in these systems, and argued in favor of a conventional $s$-wave state, possibly due to phonons.     The essential argument in these works is that magnetic impurity adatoms (Cr and Mn) are observed to create midgap bound states, whereas non-magnetic adatoms (Zn, Ag, K) did not.   There are also independent arguments proposed  in favor of conventional $s-$wave that are related to the evolution of quasiparticle interference (QPI)  peaks in a magnetic field.  In this paper, we argue, using the results of a simple phenomenological theory of impurity scattering in a multiband superconductor, that these observations cannot rule out a sign-changing $s-$wave state.

In the discussion below, we first introduce the simplest model capable of capturing the multiband effects that appear to us to be essential to understand the formation of impurity bound states in Fe-based systems, that of one hole (h) and one electron (e) band, together with a multiband pairing interaction matrix $\lambda_{ij }=N_i V_{ij}$, with $N_i$ the
Fermi level density of states and $V_{ij}$ the pairing interaction between bands $i,j=e,h$.   A nonmagnetic  impurity is then assumed to scatter within each band with amplitude $v$ and  between bands with amplitude $u$.     Even at this simple level the problem is complex, since there are several interaction and several impurity potential parameters.  Several authors have considered the symmetric model, with an $s_\pm$ configuration of  equal isotropic gaps $\Delta_e=-\Delta_h$,  and equal density of states $N_e=N_h$  for the two bands\cite{Efremov11,HKMROPP2011} as a test case that can easily be understood qualitatively.    Within the standard  $t$-matrix approximation introduced below, it was shown for this model that for general $u,v$ no  midgap bound state occurs: to find a midgap impurity state, the parameters  should be fine-tuned to very close to $u\simeq v$\cite{HKMROPP2011,BeairdVekhter}.  In a more realistic situation, with different densities of states and gaps  different on different bands, or indeed with more bands, this condition for a midgap state will be altered, but the necessity for fine-tuning will be not.  Thus it is already, for any given chemical impurity characterized by a $u,v$, very unlikely that a midgap bound state will be formed in a sign-changing gap situation\cite{HKMROPP2011,BeairdVekhter}.  The non-observation of a midgap impurity bound state is therefore, already at the level of these simple considerations, very unlikely to provide any useful information about the pair state, and certainly cannot be used to rule out $s_\pm$ pairing.     

The focus of this paper is an additional important effect that occurs in the case of incipient band $s_\pm$ pairing.  In fact, the sign-changing $s_\pm$ state in the case of an incipient band is even more robust against midgap bound state formation.    Within  the usual two-band model, this situation is considered as in Ref. \onlinecite{ChenIncipient} by simply  moving the $\Gamma$-centered hole band below the Fermi level.  We show here that in such a situation any impurity bound state is generically moved to the   gap edge $E_{min}$ and is therefore unobservable.

\begin{figure}[t]
\begin{centering}
\includegraphics[width=\columnwidth]{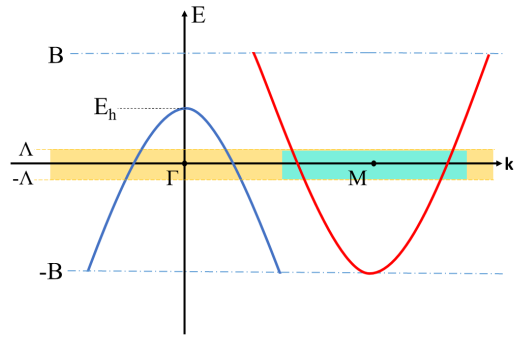}
\par\end{centering}

\caption{Band structure of the two-band model used in this paper. The yellow region denotes the range of energy over which the repulsive interband pairing interaction $V_{eh}$ is active,  whereas the range of the attractive intraband interaction $V_{ee}$ is shown in cyan.  For calculations shown here these ranges are taken to be equal.}
\label{fig:bandstructure}

\end{figure}

\emph{Impurity bound states in incipient model}. We first consider
a homogeneous two-band superconductor with the  band structure
given in Fig. 1. The upper band edge of e band is $B$, and for convenience we assume that the two bands share
the same lower edge $-B$. The maximum of the hole (h) band $E_{h}$ will be
varied continuously in the calculation. We consider a two-dimensional model where the density of states per spin
$N_{e}$ and $N_{h}$ are assumed to be constant within the band edges.
The BCS-like pairing interactions are assumed to be attractive  (negative) within   the electron band $V_{ee} <0$ and repulsive (positive) between bands, 
$V_{eh}>0$, and the BCS cutoff energies for both interactions, represented by the boundaries of the yellow region $\pm \varLambda$ in Fig. \ref{fig:bandstructure} are assumed
to be the same.  Note that these assumptions are not essential and are taken to reduce the possible number of parameters in the model.   In fact, essentially the same qualitative results will emerge if $V_{ee}$ is set equal to zero; we have assumed a weak attractive interaction in the electron band only to broaden the range of $E_h$ where a significant $T_c$ is observed.
We choose the remaining parameters to generate 
a $s_{\pm}$ gap configuration, and a robust gap on the h band in the 
incipient regime $-\varLambda<E_{h}<0$\cite{ChenIncipient}. 

In contrast 
to previous single impurity $t$-matrix calculations for particle-hole symmetric
models, two different energy scales, $\varLambda$ and $B$,
have been introduced here. In BCS theory it is normal and necessary to introduce a
pairing interaction cutoff, but the band edges typically do not play a role because particle-hole symmetry is assumed.  
In the case of an incipient
band, however, the band edge  will enter when we calculate the $t$-matrix in the
single impurity problem.   Nevertheless we will always work in the limit 
$\varLambda\ll B$, and the specific ratio between the two will not
affect the physical conclusions; we therefore set $B/\varLambda=10$ in all further 
calculations.

According to the form of the pairing interactions, the superconducting
gap can only be nonzero for states lying within the cutoff energy $\varLambda$
and only depends on band index. We will only consider $T=0$. If $-\varLambda<E_{h}<\varLambda$,
the BCS gap equation for this model reads\cite{ChenIncipient}

\begin{equation}
\Delta_{h}=-\lambda_{eh}\Delta_{e}\int_{-\varLambda}^{\varLambda}\frac{1}{2\sqrt{\varepsilon^{2}+\Delta_{e}^{2}}}d\varepsilon
\end{equation}

\begin{eqnarray}
\Delta_{e}&=&-\lambda_{ee}\Delta_{e}\int_{-\varLambda}^{\varLambda}\frac{1}{2\sqrt{\varepsilon^{2}+\Delta_{e}^{2}}}d\varepsilon \nonumber\\
&&-\lambda_{he}\Delta_{h}\int_{-\varLambda}^{E_{h}}\frac{1}{2\sqrt{\varepsilon^{2}+\Delta_{h}^{2}}}d\varepsilon
\end{eqnarray}

The gap equation in the region $\varLambda<E_{h}<B$ is obtained by
replacing $E_{h}$ in eq. (2) by $\varLambda$, thus in this situation
$\Delta_{h}$ and $\Delta_{e}$ do not depend on $E_{h}$ and are
equal to their values at $E_{h}=\varLambda$. For $-B<E_{h}<-\varLambda$,
the h band is not affected by the pairing interactions so it does not
develop a gap, $\Delta_{h}=0$; $\Delta_{e}$ does not depend on the
position of the deep h band and is equal to its value at $E_{h}=-\varLambda$;
the model  reduces to a single band $s$ wave superconductor.

Now we introduce a single static local nonmagnetic impurity. For a singlet superconductor, the single impurity $t$-matrix
$\hat{T}_{\mathbf{k},\mathbf{k}'}$, which exactly accounts for and
sums over the processes of multiple scattering off that impurity,
satisfies 

\[
\hat{T}_{\mathbf{k},\mathbf{k}'}(\omega)=\hat{U}_{\mathbf{k},\mathbf{k}'}+\sum_{\mathbf{k}^{''}}\hat{U}_{\mathbf{k},\mathbf{k}''}\hat{G}_{0}(\mathbf{k}'',\omega)\hat{T}_{\mathbf{k}'',\mathbf{k}'}(\omega)
\]

\begin{equation}
=\hat{U}_{\mathbf{k},\mathbf{k}'}+\sum_{\mathbf{k}^{''}}\hat{T}_{\mathbf{k},\mathbf{k}''}(\omega)\hat{G}_{0}(\mathbf{k}'',\omega)\hat{U}_{\mathbf{k}'',\mathbf{k}'}
\end{equation}

in which $\hat{U}_{\mathbf{k},\mathbf{k}'}$ is the scattering potential
in momentum space of the single impurity and $\hat{G}_{0}(\mathbf{k},\omega)$
is the $2\times2$ Nambu Green's function of the homogeneous system
without the impurity. In our two-band model, we assume that $\hat{U}_{\mathbf{k},\mathbf{k}'}=v\tau_{3}(u\tau_{3})$
if $\mathbf{k},\mathbf{k}'$ belongs to the same(different) band,
and with this simplification, $t$-matrix only depends on band index
and eq. (3) reduces to
 
\begin{equation}
\hat{T}_{ij}(\omega)=\hat{U}_{ij}+\sum_{l}\hat{U}_{il}\left[\sum_{\mathbf{k}^{''}}\hat{G}_{0,l}(\mathbf{k}'',\omega)\right]\hat{T}_{lj}(\omega)
\label{eq:Tij}
\end{equation}

$i,j,l=e,h$ and $\hat{U}_{ij}=v\tau_{3}(u\tau_{3})$ if $i=j(i\neq j)$.
$\hat{G}_{0,l}(\mathbf{k}'',\omega)$ denotes the Nambu Green's function.
The integrated
Green's function can be expressed as

\begin{equation}
\underset{\mathbf{k}^{''}}{\sum}\hat{G}_{0,l}(\mathbf{k}'',\omega)\equiv g_{\omega,l}\tau_{0}+g_{\Delta,l}\tau_{1}+g_{\varepsilon,l}\tau_{3}
\end{equation}


where $g_{\omega,l}=\underset{\mathbf{k}}{\sum}{\omega}/D_{\bf k}$,
$g_{\Delta,l}=\underset{\mathbf{k}}{\sum}{\Delta_{\mathbf{k},l}}/D_{\bf k}$
and $g_{\varepsilon,l}=\underset{\mathbf{k}}{\sum}{\varepsilon_{\mathbf{k},l}}/D_{\bf k}$, with $D_{\bf k}={\omega^{2}-\Delta_{\mathbf{k},l}^{2}-\varepsilon_{\mathbf{k},l}^{2}}$ are the Nambu components of the local Green's function. 
In the usual particle-hole symmetric
models, all bands cross Fermi level and the band edges are assumed
to be much larger than other energy scales in the problem; 
 the positions of the band edges are therefore irrelevant, and $g_{\varepsilon,l}$
 vanishes.   However, for an incipient  band which
is close to the Fermi level, $g_{\varepsilon,l}$
is nonzero and becomes important since its magnitude is generally
much larger than $g_{\omega,l}$ and $g_{\Delta,l}$ for $\omega\sim\Delta$. 
If we assume a constant density of states for an incipient 2D band $l$,
$g_{\varepsilon,l}$ is logarithmically divergent at large $|\varepsilon|$,
thus it is necessary to include a higher energy scale, e. g. the band
edge $B$ as in our model, to truncate the integration and generate
a physical result for $g_{\varepsilon,l}$.

For a conventional $s$ wave superconductor, the energy interval centered
at the  Fermi level in which total density of states is zero is determined
by the lowest quasiparticle energy $E_{\mathbf{k}}=\sqrt{\Delta_{\mathbf{k}}^{2}+\varepsilon_{\mathbf{k}}^{2}}$
of the system. If we denote this energy by $E_{min}$, that energy
interval is $(-E_{min},E_{min})$. Outside this interval is the continuous
part of the density of states  observed in tunneling
experiments. In our model, if $E_{h}>0$, $E_{min}=\min(|\Delta_{h}|,|\Delta_{e}|)$;
if $E_{h}<0$, $E_{min}=\min(\sqrt{E_{h}^{2}+\Delta_{h}^{2}},|\Delta_{e}|)$.
The positions of the poles on real axis in the interval $(-E_{min},E_{min})$
of the $t$-matrix $\hat{T}_{\mathbf{k},\mathbf{k}'}$, considered as a function of
$\omega$, are the ``in-gap'' impurity bound state energies. The $t$-matrix $\hat{T}_{ij}$ in our model is solved
 from Eq. \ref{eq:Tij}.
The denominator $D(\omega)$ of $\hat{T}_{ij}$ (independent
of $i,j$) is found as 
\begin{widetext}
\begin{eqnarray}
D(\omega)&=&(1-2g_{\varepsilon,h}v+Av^{2})(1-2g_{\varepsilon,e}v+Bv^{2})+\\
&&2\left[g_{\Delta,e}g_{\Delta,h}-g_{\omega,e}g_{\omega,h}-(g_{\varepsilon,h}-Av)(g_{\varepsilon,e}-Bv)\right]u^{2}+ABu^{4} \nonumber
\end{eqnarray}
\end{widetext}
in which $A=(g_{\Delta,h}^{2}+g_{\varepsilon,h}^{2}-g_{\omega,h}^{2})$, $B=(g_{\Delta,e}^{2}+g_{\varepsilon,e}^{2}-g_{\omega,e}^{2})$.

\begin{figure}[h]
\noindent \centering{}\includegraphics[width=0.9\columnwidth]{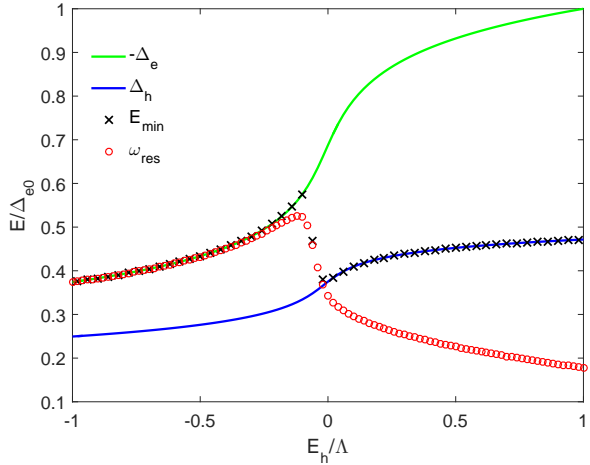}\caption{Impurity bound state energy and gap values  for parameters: $N_{e}V_{ee}=-0.3$, $N_{h}V_{eh}=0.2$, $N_{e}V_{eh}=0.2$,
$N_{h}v=N_e v=2$, $N_{h}u=N_e u =2$. $\Delta_{e0}$ is $|\Delta_e|$ at $E_h=\Lambda$. The green and blue solid lines are $-\Delta_{e}$,
$\Delta_{h}$ vs. $E_{h}$ respectively. The black crosses represent
$E_{min}$, the edge of continuous part of LDOS(or total DoS) on positive
$\omega$ as a function of $E_{h}$. The red circles are positive
single impurity bound state energy.}
\label{fig:boundstate1}
\end{figure}

Our goal now is to show that, even under circumstances where a nonmagnetic  impurity might give rise to midgap states, such states (and their pairbreaking capacity) rapidly disappear when the hole band is moved below the Fermi surface.   
Since $D$ as a function of real $\omega$
is even, we only need to find the root of $D=0$ for $0\leqslant\omega<E_{min}$.   As discussed in several papers, the most likely situation for a midgap impurity state in an $s_\pm$ state is when inter- and interband scattering rates are comparable, $u\approx v$.  We therefore assume this condition,   and indeed find a midgap state in the symmetric band limit, which persists while the hole band remains at the Fermi level, as shown in Fig. \ref{fig:boundstate1}.  As $E_h$ is lowered in the plot, the self-consistently calculated gaps $\Delta_e$ and $\Delta_h$ are seen to decrease, and as the Lifshitz transition is crossed fall  more rapidly.  The intraband attraction assumed in the electron band for this particular case supports the gap significantly even when $E_h<0$, so that we can observe the impurity bound state behavior over a larger range.  In the region of the Lifshitz transtion, the bound state is seen to pass through the  hole gap energy. In our model, the in-gap bound states never move into the quasiparticle continuum, but tail onto the electron gap edge.

\begin{figure}[h]
\noindent \centering{}\includegraphics[width=0.49\columnwidth]{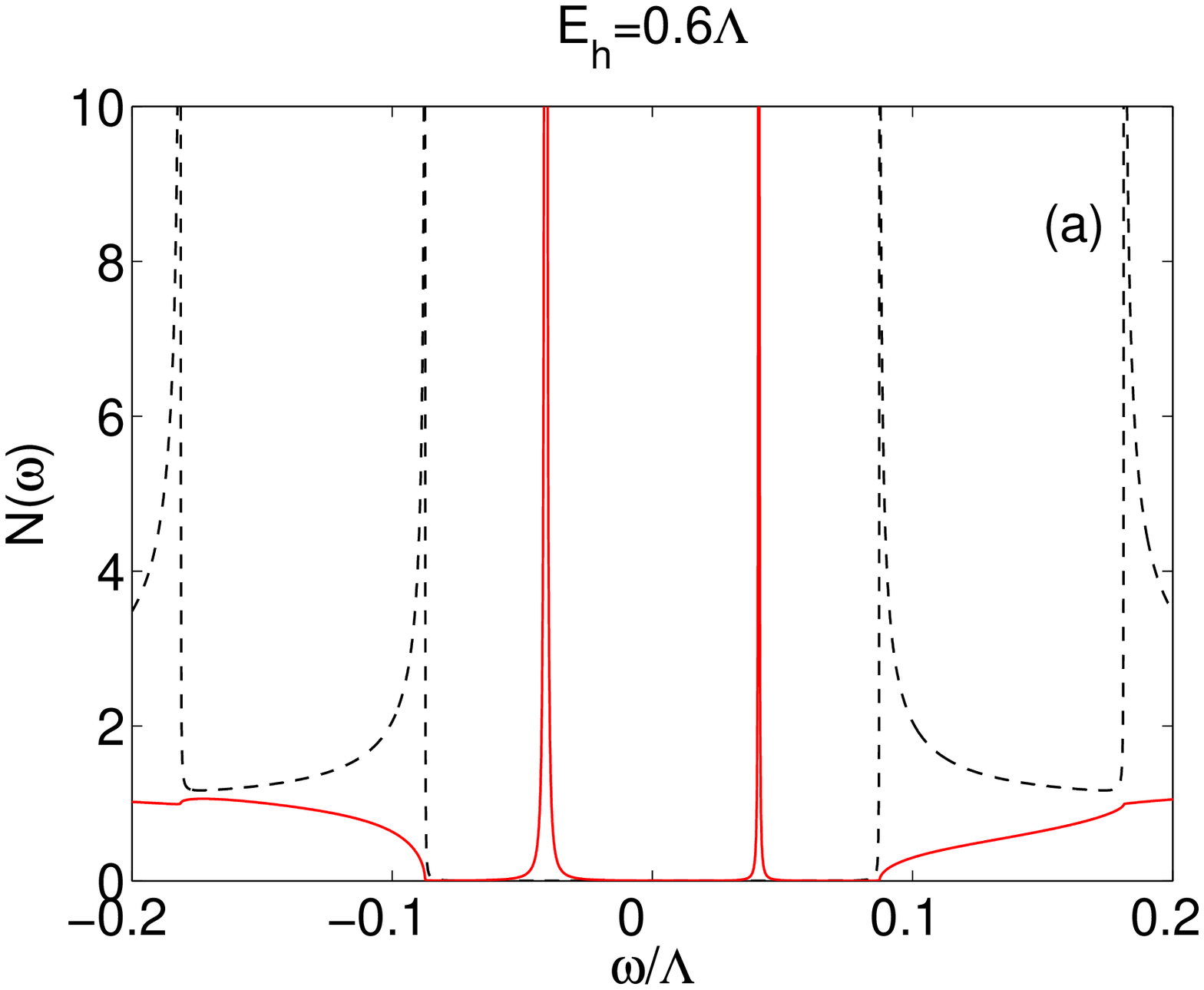}\includegraphics[width=0.49\columnwidth]{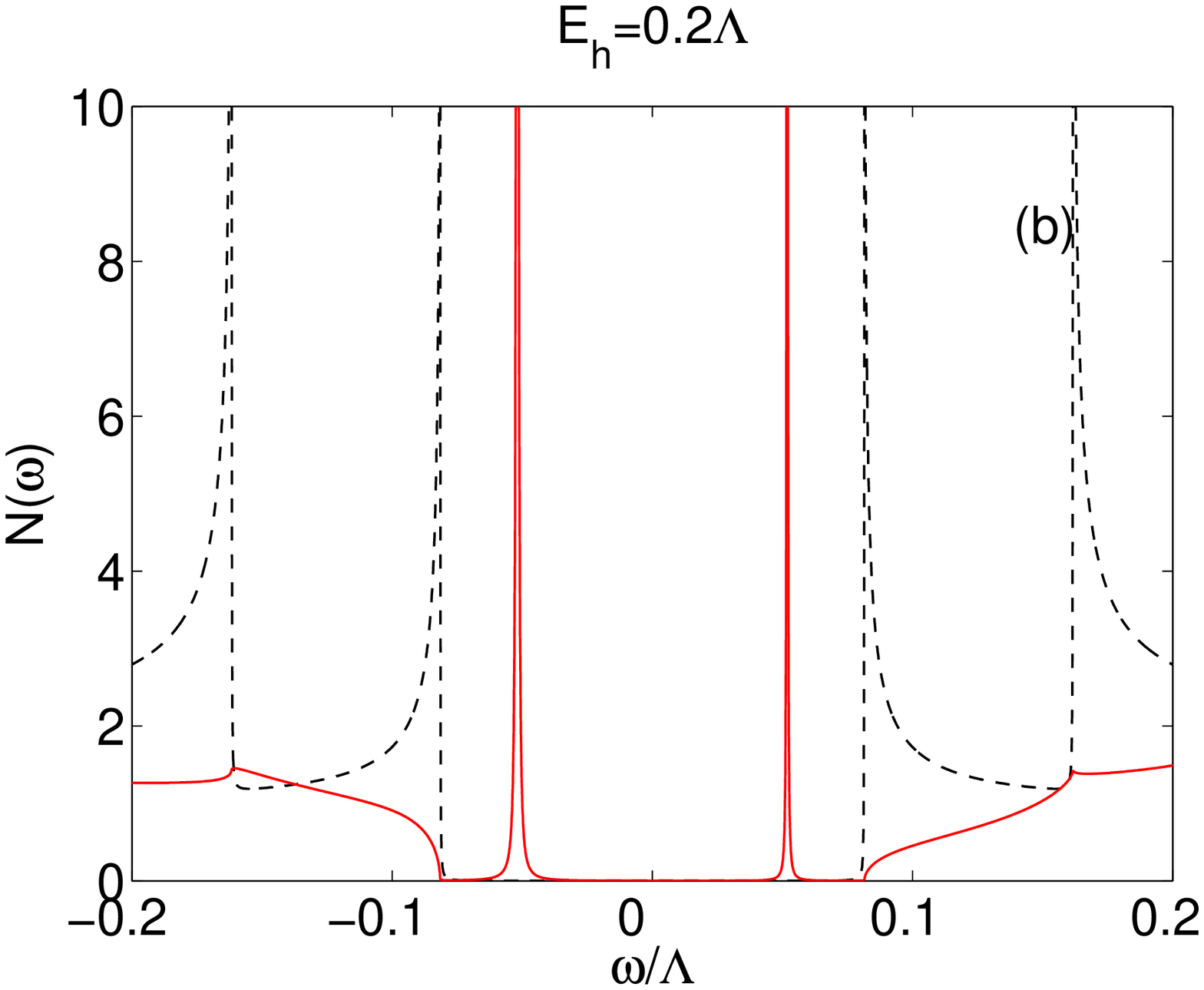}
\includegraphics[width=0.49\columnwidth]{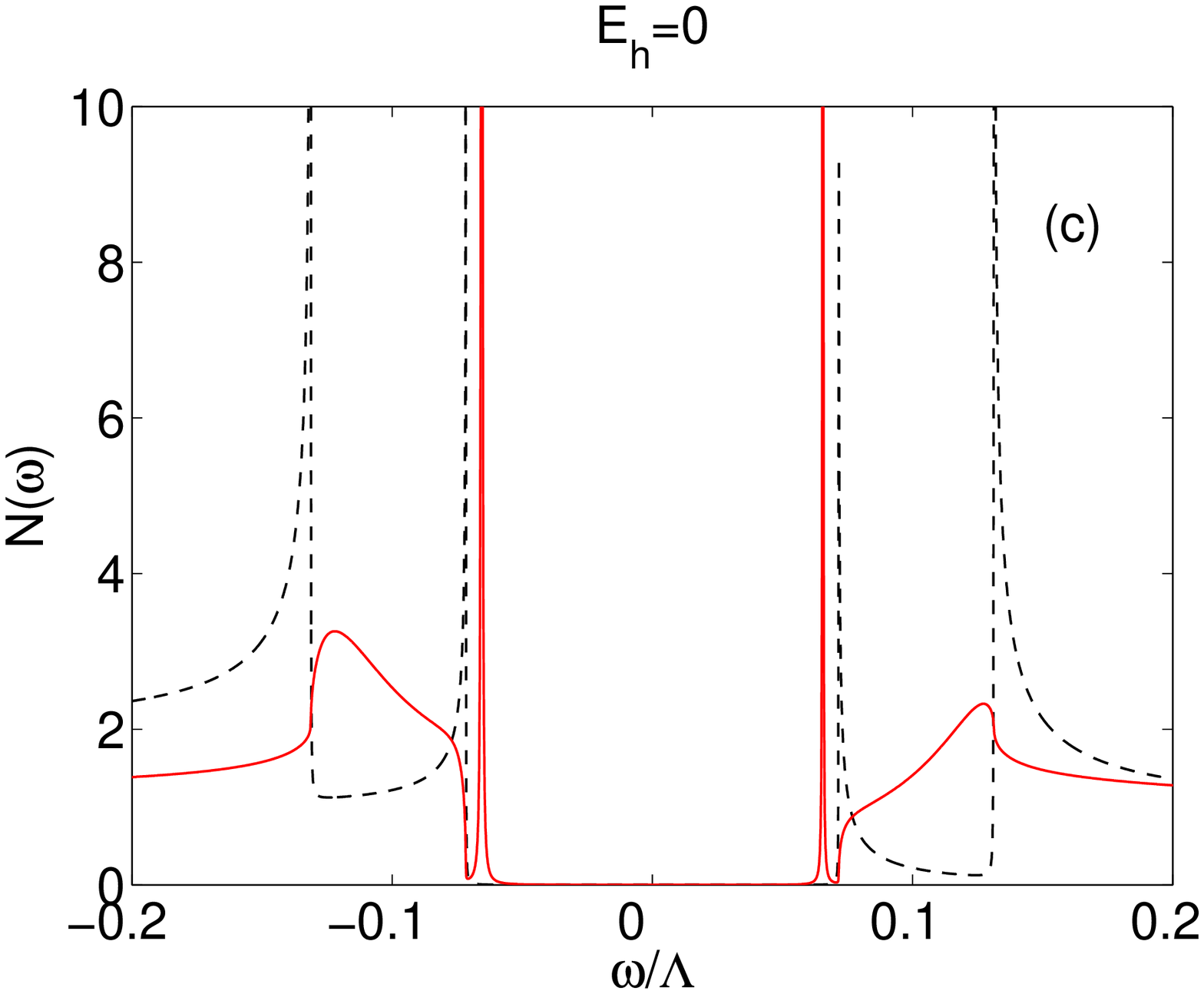}\includegraphics[width=0.49\columnwidth]{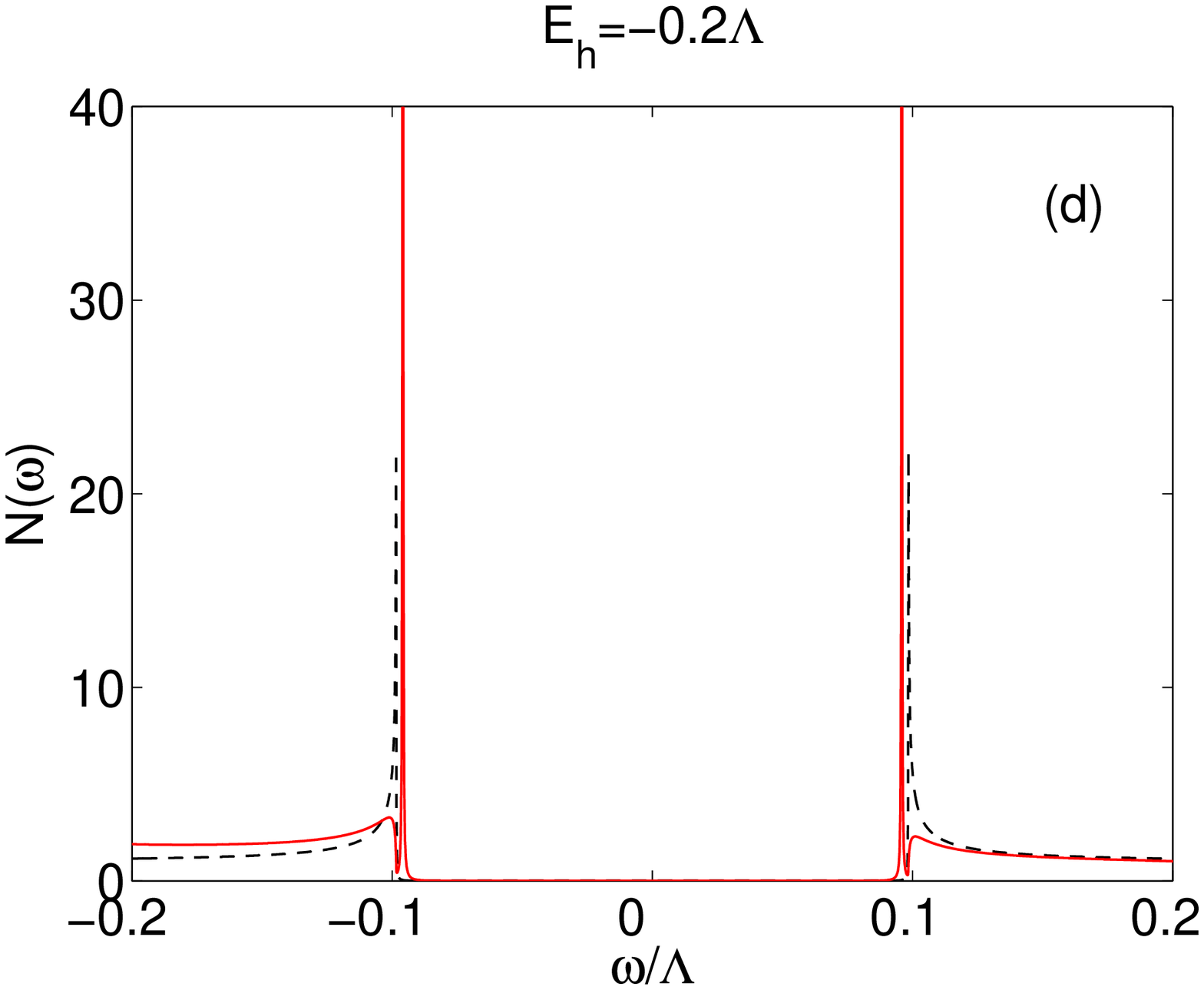}

\caption{Local density of states at the impurity site (red curve) and in bulk (black dashed curve) for various values of $E_h$ for the case, when the gap on electron band is larger than the gap on incipient hole band. }
\label{fig:LDOS1}
\end{figure}

The evolution of the corresponding local density of states  is now shown in Fig. \ref{fig:LDOS1}.  For impurity potentials that place the bound states in the gap region of an $s_\pm$ superconductor (Fig. \ref{fig:LDOS1}(a)), as the hole band is moved down, these bound states move closer to the lower gap edge, which is determined by the hole band in this figure.  Eventually, the Lifshitz transition is reached (Fig. \ref{fig:LDOS1}(c)),  the coherence peak features  at $\pm\Delta_h$ is replaced by weak particle-hole asymmetric features at  $\pm\sqrt{\Delta_h^2+E_h^2}~$\cite{Koshelev}, and the position of the bound state is seen to saturate at the electron gap edge (Fig. \ref{fig:LDOS1}(d)).  In the true incipient case,  the bound state is effectively invisible.

\begin{figure}[h]
\noindent \begin{centering}
\includegraphics[width=0.9\columnwidth]{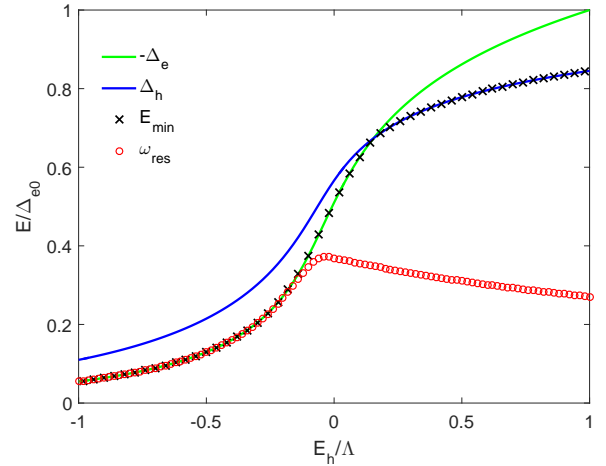}
\par\end{centering}

\caption{Same as Fig. \ref{fig:boundstate1}, but for parameters: $N_{e}V_{ee}=-0.2$, $N_{h}V_{eh}=0.3$, $N_{e}V_{eh}=0.4$,
$N_{h}v=2$, $N_{h}u=2$.}
\label{fig:boundstate2}
\end{figure}

We now consider parameters such that the hole band gap in the symmetric case $\Delta_h$ is comparable to the electron band gap $\Delta_e$, and ask how the gaps and  bound state evolve.  This is shown in Fig. \ref{fig:boundstate2}, where again we begin at large $E_h$ with a well-defined midgap impurity state.  As the hole band is lowered, the $\Delta_h$ quickly  becomes the largest gap in the system, as also found in Ref. \onlinecite{ChenIncipient}, but again below the Lifshitz transition the bound state energy is pinned at the electron gap  edge.

\begin{figure}[h]
\noindent \centering{}\includegraphics[width=0.49\columnwidth]{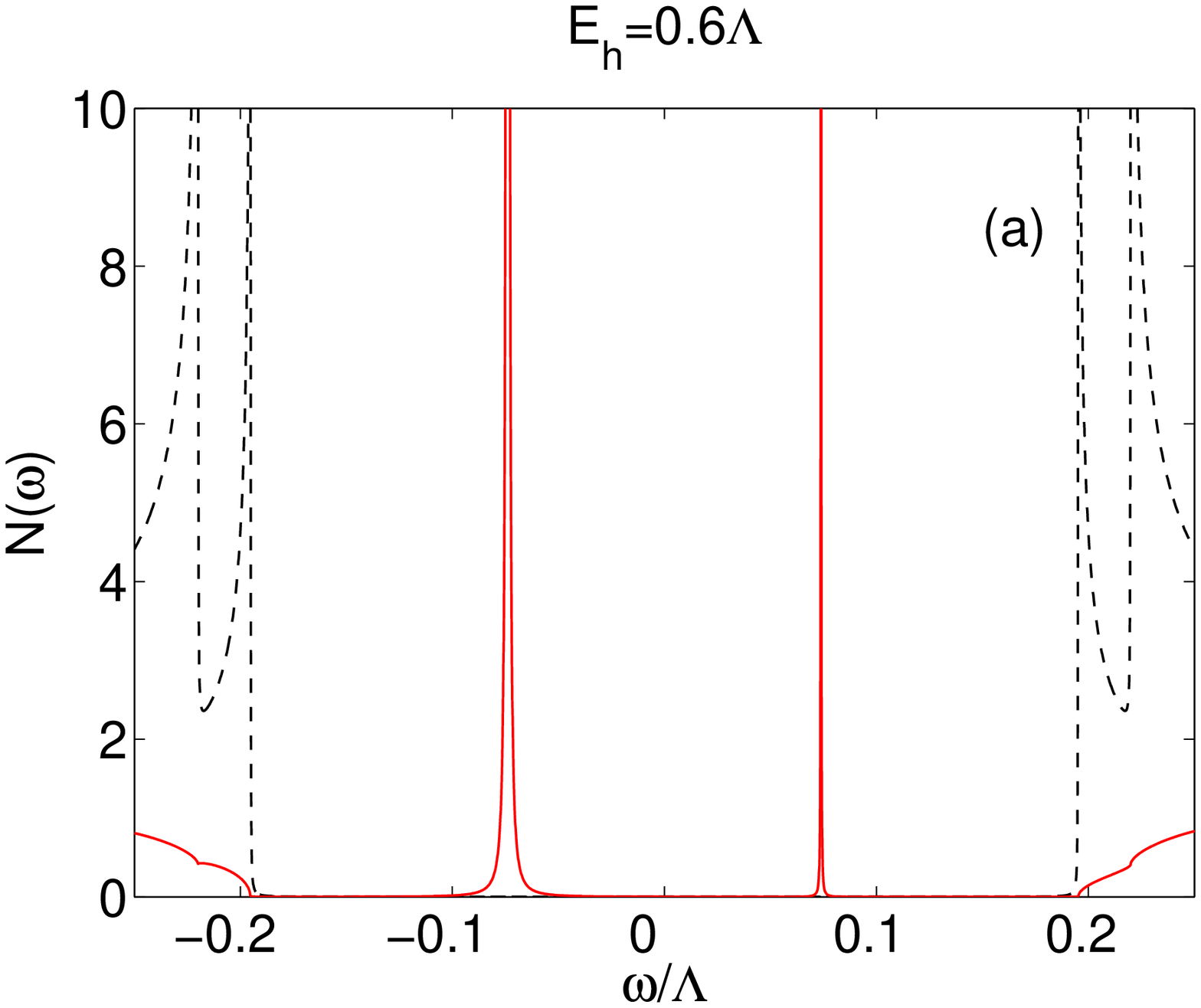}\includegraphics[width=0.49\columnwidth]{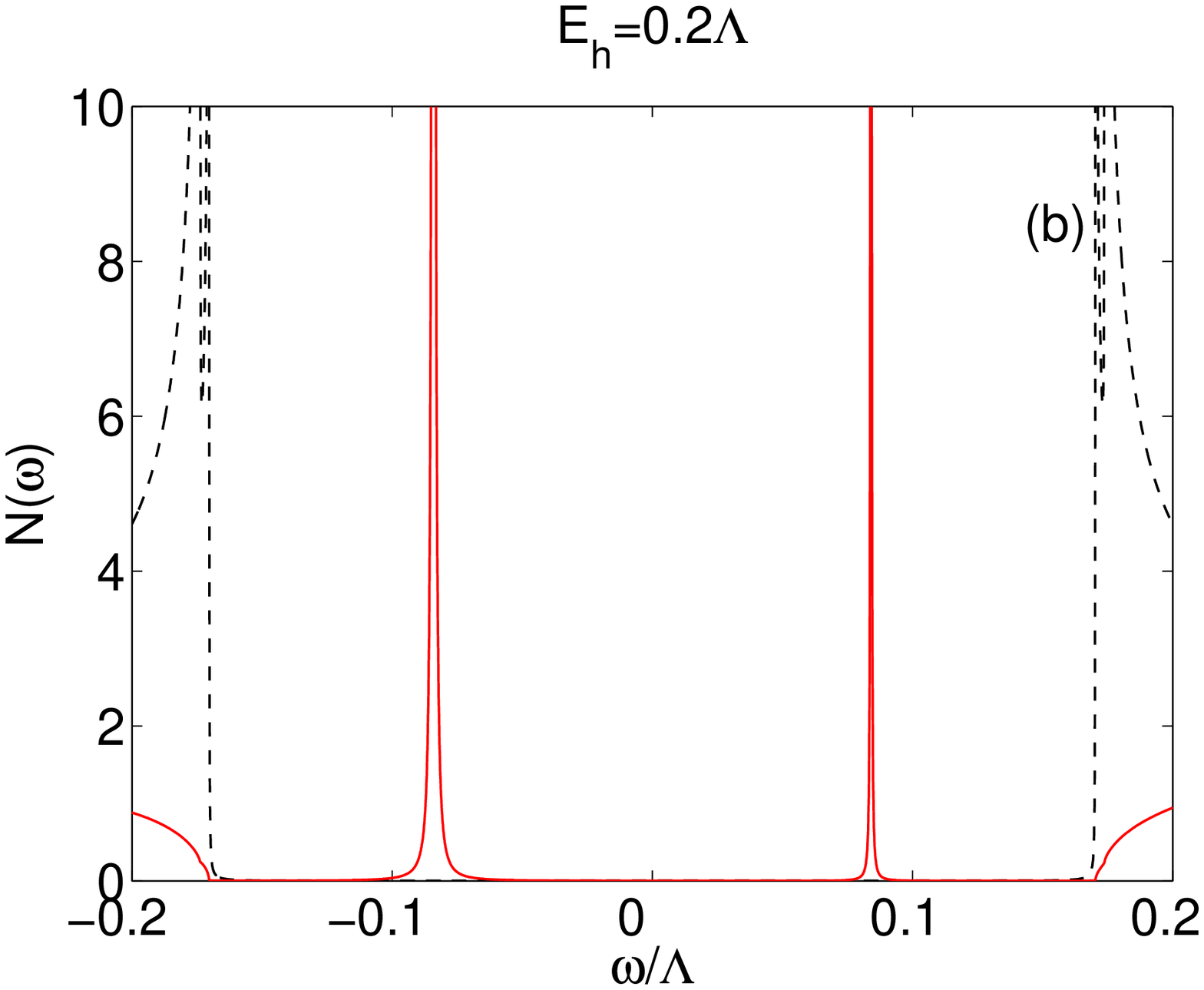}
\includegraphics[width=0.49\columnwidth]{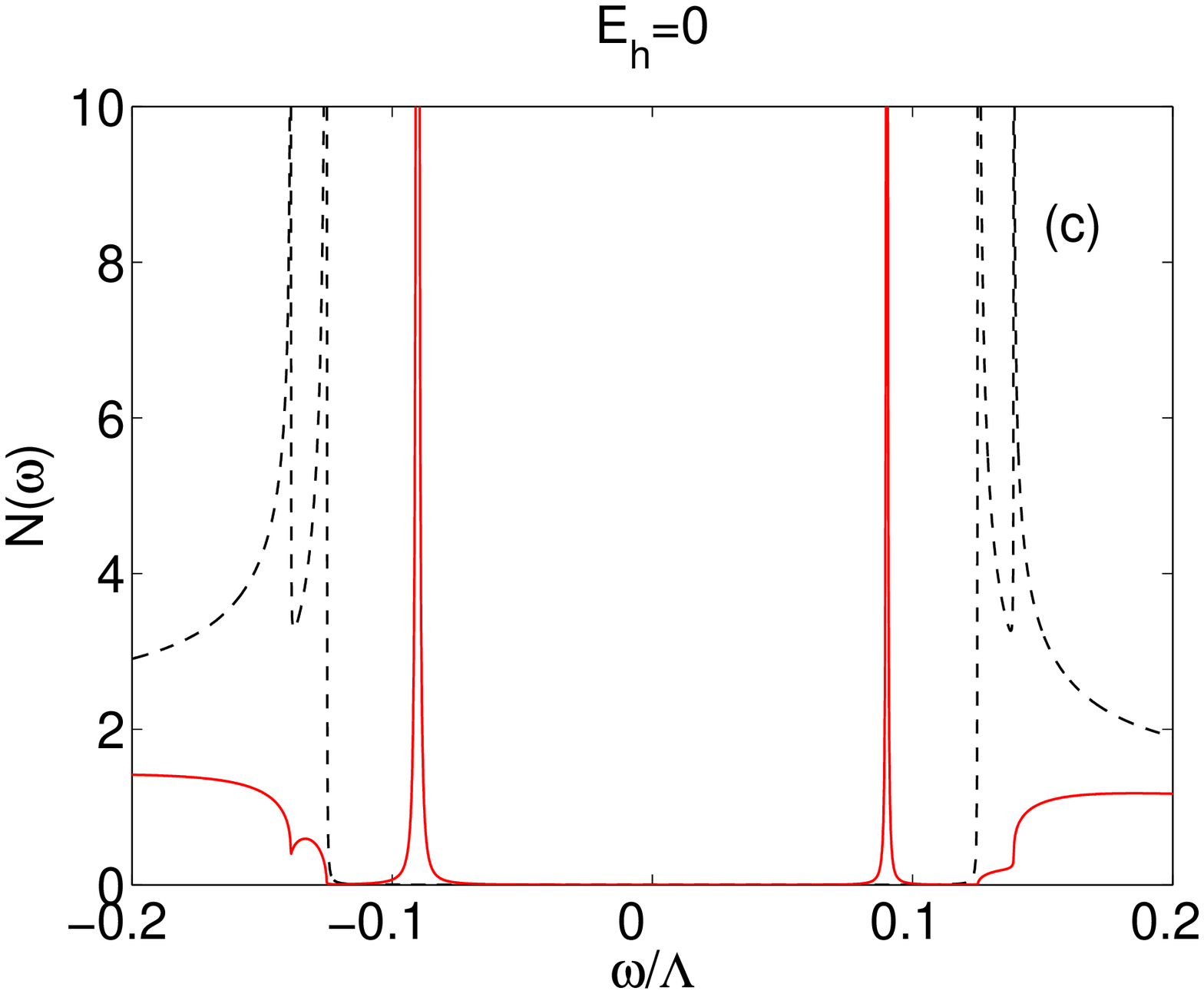}\includegraphics[width=0.49\columnwidth]{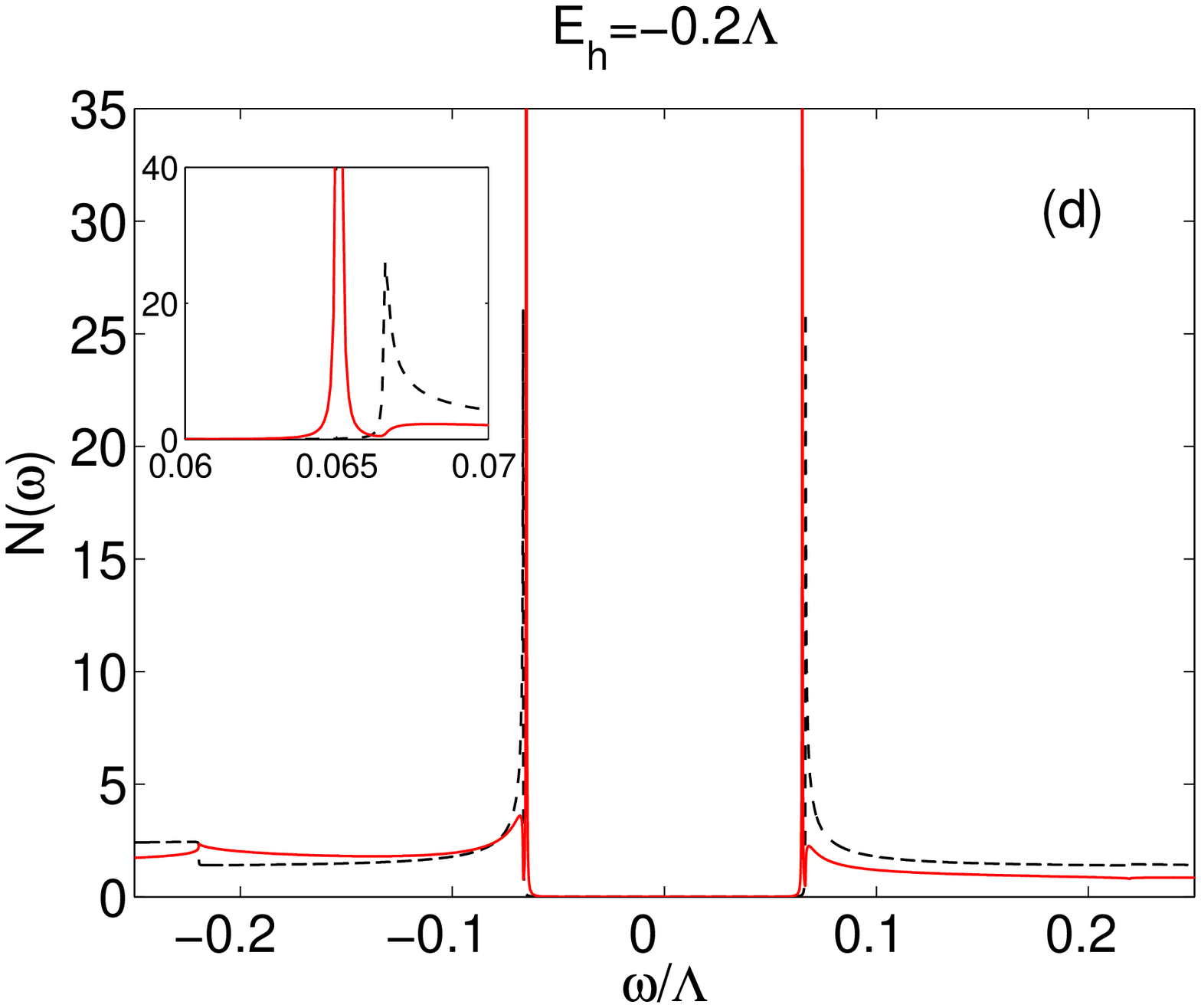}
\caption{Local density of states at the impurity site (red curve) and in bulk (black dashed curve) for various values of $E_h$ for the parameters given in Fig. \ref{fig:boundstate2}}.
\label{fig:LDOS2}
\end{figure}

The evolution of the LDOS for the case exhibited in Fig. \ref{fig:boundstate2} is now depicted in Fig. \ref{fig:LDOS2}.  Again, the bound states are seen to  lie exactly at the electron band gap edge when the hole band moves below the Fermi level.


We found from our calculation that an  impurity bound state 
exists for any $-B<E_{h}<B$ irrespective of pairing interactions
and impurity scattering potential. However, if the bound state energy is not already
close to $E_{min}$, the edge of the continuum, at $E_{h}=B$, it approaches $E_{min}$ rapidly
as the top of the h band $E_{h}$ goes below the Fermi level.  Note that if the bound state energy initially lies  close to the lower gap edge at $E_h=B$, it is found to always stay near $E_{min}$ as the h band is moved down. If $E_{h}<0$,
as long as $|E_{h}|$ is larger than $|\Delta_{e}|$, $E_{min}$=$|\Delta_{e}|$,
thus in this situation the bound state energy is found to be near
the gap edge of the electron band.

To check whether our conclusions are robust against the inclusion of further bands in the problem, we have examined other multiband cases.  For the situation where superconductivity is supported primarily by repulsive interactions between states at the Fermi level (labeled ii(a)  in Ref. \onlinecite{ChenIncipient}), the incipient band does not play an essential role.  Any nonmagnetic bound states formed in such a situation (subject to the caveats discussed in the introduction) are robust against the introduction of an incipient band.  

 For the case of a magnetic impurity,  however, even in the situation without Fermi surface hole pockets discussed here, 
 one finds the usual Yu-Shiba-Russinov type bound states, as observed in experiment\cite{Fan_etal_NatPhys2015,Yan_etal_aXv2015}.

\begin{figure}[h]
\noindent \begin{centering}
\includegraphics[width=0.9\columnwidth]{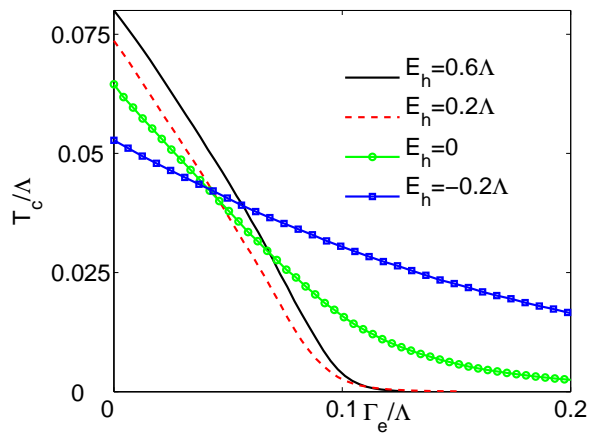}
\par\end{centering}
\caption{Critical temperature calculated from the BCS gap equation for the system parameters of Fig. \ref{fig:boundstate1} vs. electron band scattering rate $\Gamma_e$.  Both scales are given in energy unit of $\varLambda$. }
\label{fig:Tcsupp1}
\end{figure}

\begin{figure}[h]
\noindent \begin{centering}
\includegraphics[width=0.9\columnwidth]{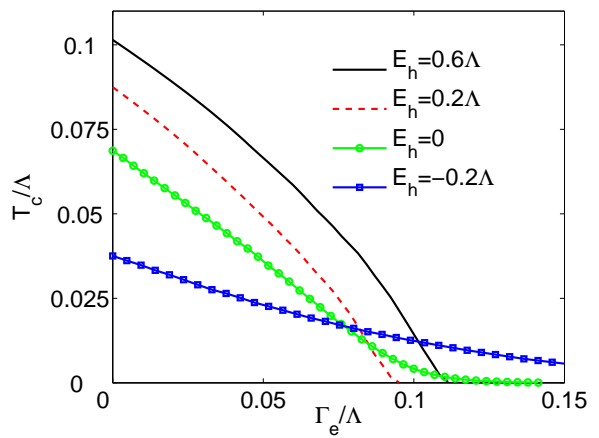}
\par\end{centering}
\caption{Same as Fig. \ref{fig:Tcsupp1} but for the parameters of Fig. \ref{fig:boundstate2}.}
\label{fig:Tcsupp2}
\end{figure}

\emph{$T_{c}$ suppression.}
Here we study similar representative cases as in the previous section, but for finite nonmagnetic disorder.  In general, the lack of impurity bound state formation in the single impurity case signals  weak pairbreaking overall, and this is the case with incipient band pairing as well.  In Fig. \ref{fig:Tcsupp1}, we show that in the pure sample, $T_c$ is suppressed as usual as $E_h$ is lowered (note that this suppression is not particularly rapid due to the assumption of a finite intraband $V_{ee}$, which is employed simply to spread out the range of interesting $E_h$, as discussed above).  
Within the t-matrix approximation,  the self energy is given by
\begin{equation}
\hat{\Sigma}_{e/h}= n_{imp} \hat{T}_{ii}.
\end{equation}
Note, averaging over random configuration of impurities restores the translational invariance  
of the system  and makes the self-energy diagonal in the band basis.
However,
it has $\tau_0$, $\tau_{1}$ and $\tau_3$ components in the Nambu basis. For simplicity, we ignore
the $\tau_3$ component of self-energy, which mainly contributes to renormalization of the Lifshitz point
by changing the chemical potential.
The degree of scattering can therefore be parametrized, e.g. by the normal state scattering rate in the zero temperature limit $\Gamma_e = -{1\over 2}{\rm Im Tr} \hat{\Sigma}_e$. In general, gaps on the electron and hole band are determined by solving gap equations,
\begin{eqnarray}
\Delta_{i}&=& 2 T \sum_{\omega_n > 0,j}^{\varLambda}  \frac{ -\lambda_{ij} \tilde{\Delta}_j}{Q_j} \left[ \tan^{-1}\left( \frac{B}{Q_j}\right)+ \tan^{-1}\left( \frac{E_j}{Q_j}\right) \right],  \nonumber\\
\tilde{\omega}_{nj} &=& \omega_n + \frac{1}{2} {\rm Tr}\, \hat \Sigma_j,~~~~
\tilde{\Delta}_{j} = \Delta_j + \frac{1}{2} {\rm Tr} \left [\tau_1 \hat \Sigma_j \right],
\end{eqnarray}
where $Q_j =\sqrt{\tilde{\omega_{nj}}^2+\tilde{\Delta}^2_j}$ and $E_j$ is $B$ for the electron band and $E_h$ for the hole band. $T_c$, is now determined in the usual way by linearizing the gap equations and self-energy equations to first order in $\Delta_{e/h}$.   As expected, the effect of finite disorder as $E_h$ is lowered is to decrease the pairbreaking rate, as seen by the initial slope of $T_c$ decreasing and the persistence of superconductivity to higher scattering rates.  For completeness, we exhibit in Fig. \ref{fig:Tcsupp2} the same phenomenon for the case where the hole gap is larger in the incipient band limit. 

{ Based on the results shown in Figs. \ref{fig:Tcsupp1} and \ref{fig:Tcsupp2} we see that even though the Fermi surface is  only comprised of a single band, there is a $T_c$ suppression with increasing disorder ($\Gamma$), whose rate decreases as the ratio $\Gamma/|E_h|$ decreases.  This result is counter to a naive application of Anderson's theorem for an isotropic 1-band SC, and  indicates the vital role of scattering from the incipient band. Only in the limit of a deep incipient band ($\Gamma/|E_h|\rightarrow 0$) does one restore the Anderson criteria as the slope of the $T_c$ suppression $\rightarrow 0$.}

\emph{Quasiparticle Interference.}
Reference \onlinecite{Yan_etal_aXv2015} argued that
while a QPI peak corresponding to $\Gamma-M$ scattering is observed in (Li$_{0.8}$Fe$_{0.2}$)OHFeSe, indicating the existence of significant large-q scattering processes near the Fermi level involving the incipient band,  a magnetic field does not distinguish between small-q and large-q peaks.  As discussed in Ref. \onlinecite{PJH_QPI15}, however, there is no theoretical justification for the commonly held assumption that a magnetic field can lead to suppression of QPI peaks for wave vectors that connect gaps of different sign.  The lack of observation of such suppression cannot therefore be used as an argument against $s_\pm$ pairing either.  

\emph{Conclusions.}
The Fe-chalcogenide superconductors whose Fermi surface is lacking the $\Gamma$-centered hole pockets characteristic of the Fe-pnictide systems have been the subject of intensive debate.  Recently, several papers have put forward evidence from STM measurements arguing    that because  nonmagnetic impurity bound state features are not observed, non-sign changing pairing states are realized.  Here
we have shown, on the contrary, that if the incipient band near the $\Gamma$ point plays an essential role in the pairing, which earlier work has shown can be the case, nonmagnetic impurity  bound states are essentially unobservable due to their location at the gap edge.
We have also pointed out  that the counterargument is not correct: observation of nonmagnetic midgap impurity bound states does not rule out  pairing on the incipient band, but rather simply indicates the existence of more robust pairing taking place in the states at the Fermi level.  

In addition, we have examined the effect of finite disorder on systems where  incipient bands play an essential role in the pairing.  Consistent with the conclusions regarding bound state formation, we find that the $T_c$ suppression rate due to disorder is substantially suppressed in absolute units.

Our work has important implications for the discussion of the possible ground states of the Fe-based superconductors without $\Gamma$ centered pockets at the Fermi level, and leaves open the possibility that ``conventional'' $s_\pm$ states involving sign change of the superconducting gap between hole and electron pockets may still be realized.

\emph{Acknowledgements.}  The authors are grateful for useful discussions with A. Linscheid, S. Johnston, and Y. Wang.  XC, SM and PJH were supported by NSF-DMR-1005625. VM was supported by the Laboratory Directed
Research and Development Program of Oak Ridge National Laboratory, managed by
UT-Battelle, LLC, for the U. S. Department of Energy. 
\bibliographystyle{apsrev4-1}

\end{document}